\documentclass[twocolumn,showpacs,preprintnumbers,amsmath,amssymb]{revtex4}
\usepackage[dvips]{graphicx}
\usepackage{longtable}
\usepackage{color}

\def\cor#1{#1}

\def\ket#1{  \left\vert  #1   \right\rangle   }
\def\bra#1{  \left\langle  #1   \right\vert   }
\def\etal{\textit{et al.}}

\begin{document}

\title{Evolution equation of entanglement for multi-qubit systems}

\author{Michael Siomau}
\email{siomau@physi.uni-heidelberg.de}
\affiliation{Max-Planck-Institut f\"{u}r Kernphysik, Postfach
             103980, D-69029 Heidelberg, Germany}
\affiliation{Physikalisches Institut, Heidelberg Universit\"{a}t,
D-69120 Heidelberg, Germany}

\author{Stephan Fritzsche}
\affiliation{Department of Physical Sciences, P.O.~Box 3000,
             Fin-90014 University of Oulu, Finland}
\affiliation{GSI Helmholtzzentrum f\"{u}r Schwerionenforschung,
D-64291 Darmstadt, Germany}

\begin{abstract}
We discuss entanglement evolution of a multi-qubit system when one
of its qubits is subjected to a general noisy channel. For such a
system, an evolution equation of entanglement for a lower bound for
multi-qubit concurrence is derived. Using this evolution equation,
the entanglement dynamics of an initially mixed three-qubit state
composed of a GHZ and a W state is analyzed if one of the qubits is
affected by a phase, an amplitude or a generalized amplitude damping
channel.
\end{abstract}

\pacs{03.67.Mn, 03.65.Yz.}

\maketitle

\section{\label{sec:1} Introduction}

Entanglement is an exclusive nonclassical resource of quantum
mechanics with promising applications in communication, cryptography
and computing \cite{Nielsen:00}. \cor{In many of these
applications}, this resource is often required to be distributed
between different partners; a request that implies that at least one
physical (sub-)system is to be transmitted through a communication
channel. In general, such a coupling of a quantum system to some
environmental channel leads to decoherence of the system and usually
to \cor{some} loss of entanglement before the state of the system
can be further utilized. Therefore, it is of great practical
importance to investigate the decoherence and time evolution of
entanglement for quantum systems that undergo the action of noisy
channels.

Typically, the time evolution of entanglement of a system is deduced
from studying its state evolution under the influence of decoherence
\cite{Carvalho:04,Mintert:05,Yu:06,Siomau:10}. Following this line,
we have recently analyzed \cor{for example} the entanglement
dynamics of initially prepared pure three-qubit GHZ and W states if
transmitted through one of the Pauli channels $\sigma_z, \,
\sigma_x,\, \sigma_y$ or a depolarizing channel \cite{Siomau:10}. In
this work, we employed the analytical solutions of the master
equation as obtained by Jung \etal{} \cite{Jung:08} and a lower
bound for the three-qubit concurrence \cite{Li(b):09} in order to
quantify the time-dependent entanglement of the mixed three-qubit
states.

Instead of making explicit use of the state evolution for the
analysis of the entanglement dynamics of a given system, Konrad
\etal{} \cite{Konrad:08} recently derived an evolution equation of
entanglement for a \textit{two-qubit} system that provides a direct
relationship between the initial and the final entanglement of the
system when one of its qubits is subjected to an arbitrary noise.
Subsequently, Li \etal{} \cite{Li(d):09} derived a generalized
evolution equation of entanglement for a (finite dimensional)
\textit{bipartite} system, if initially prepared in a pure state and
affected by an arbitrary noisy channel. Recently, moreover, this
latter result \cite{Li(d):09} has been extended to the case of an
initial mixed state of a bipartite system \cite{Liu:09,Yu:08}.

In this work, we suggest an evolution equation of entanglement for a
\textit{multi-qubit} system when one of its qubits undergoes the
action of an arbitrary channel, \cor{which is given by a completely
positive (non-)trace-preserving map}. Since there is no an
analytically computable measure of entanglement for mixed
multi-qubit states \cite{Horodecki:09}, we shall make use of the
lower bound by Li \etal{} \cite{Li(b):09} for the multi-qubit
concurrence. In fact, this lower bound was proven to satisfy the
general requirements \cite{Mintert:05} of an entanglement measure
\cite{Li(b):09,Ou:08}. Therefore, the evolution equation derived
below \cor{is capable to} characterize the entanglement dynamics of
multi-qubit states representing a lower bound for an actual
currently unavailable entanglement evolution. As example, we shall
discuss in details the time evolution of entanglement of an
initially mixed three-qubit state composed of a GHZ and a W state
\cite{Lohmayer:06}
\begin{equation}
 \label{the-mixed-state}
 \rho(p) = p \ket{GHZ}\bra{GHZ} + (1-p) \ket{W}\bra{W} \, ,
\end{equation}
if one of the qubits is affected by a phase or an amplitude damping
channel. For a generalized amplitude damping channel, moreover, we
also show the sudden death of entanglement \cite{Yu:06} for this
mixed three-qubit state.

\cor{The paper is organized as follows.} In the next section, we
first recall how the entanglement of a multi-qubit system can be
quantified \cor{either by means of the $N$-qubit concurrence
\cite{Mintert:05,Horodecki:09} or in terms of a lower bound to this
measure as suggested by Li \etal{} \cite{Li(b):09}.} This lower
bound was constructed in such a way that it includes only bipartite
concurrences according to some ``bi-partite'' cuts of the
multi-qubit system. In Section.~\ref{subsec:2.2}, we display and
discuss the evolution equation of bipartite concurrence as obtained
by Li \etal{} \cite{Li(d):09}. Based on their results for the
bipartite concurrence, \cor{we then derive an evolution equation of
the lower bound to the concurrence for a three-qubit system in
Section II.C, if one of its qubits is affected by an arbitrary noisy
channel and if we start with an initially pure state. In
Section~\ref{subsec:2.4}, we shall discuss also possible extensions
of this equation to the cases of N qubits, initially mixed states of
three qubits, and if a system is affected by many-sided noisy
channels, i.e.~when several of its qubits undergo simultaneously the
action of some local noise. In Section~\ref{sec:3}, we later analyze
in detail the entanglement dynamics of the initially three-qubit
mixed state (\ref{the-mixed-state}) for three realistic noise
models, namely, a phase damping, an amplitude damping and a
generalized amplitude damping noise. Finally, a conclusion is drawn
in Section~\ref{sec:4}.}

\section{\label{sec:2} Evolution equation of the lower bound for
                       multi-qubit concurrence}

Until the present, it has been found difficult to quantify the
entanglement of mixed many-partite states, and no general solution
is known \cite{Horodecki:09} apart from Wootter's (two-qubit)
\textit{concurrence} \cite{Wootters:98}. This concurrence provides
indeed a very powerful measure of entanglement but is just suitable
for two-qubit systems. Various extensions of Wootter's concurrence
have been worked out over the years, and especially for mixed
bipartite states, if the dimensions of the associated Hilbert
(sub-)spaces are larger than two
\cite{Horodecki:09,Coffman:00,Rungta:01}. Nonetheless, a full
generalization of the concurrence towards mixed many-partite states
has remained a challenge until now and will require further studies
\cite{Horodecki:09}. In practice, however, one is mostly interested
in the minimum (nontrivial) amount of entanglement \cite{Guehne:09}
which is preserved in a mixed state of some system when coupled to
its environment, for instance, due to transmission of one or several
of its subsystems through a communication channel. To this end,
various analytically computable lower bounds to concurrence were
suggested recently \cite{Mintert:05,Chen:05,Li(b):09}. In the next
section, we shall exploit the lower bound to the multi-qubit
concurrence as suggested by Li \etal{} \cite{Li(b):09}. As mentioned
before, this bound is based on the bipartite concurrences of the
multi-qubit system with regard to some bipartite cuts. This
``bi-partitioning'' enables us eventually to construct an evolution
equation for multi-qubit systems which is based on the evolution
equation for bipartite systems \cite{Li(d):09}.

\subsection{\label{subsec:2.1} A lower bound for $N$-qubit concurrence}

For a given \textit{pure} N-qubit state $\ket{\psi}$, the
concurrence can be written as \cite{Albeverio:01,Li(b):09}
\begin{equation}
\label{c-pure}
C_N(\ket{\psi}) = \sqrt{1 - \frac{1}{N} \sum_{i=1}^N {\rm Tr}\,
\rho_i^2 } \, ,
\end{equation}
and where the $\rho_i = {\rm Tr} \ket{\psi}\bra{\psi}$ denote the
reduced density matrix of the $i$-th qubit which is obtained by
tracing out the remaining $N-1$ qubits. Since any \textit{mixed}
state can be expressed also as a convex sum of some pure states $\{
\ket{\psi_i} \}$: $\rho = \sum_i \, p_i \ket{\psi_i}\bra{\psi_i}$,
the definition (\ref{c-pure}) for the concurrence for pure N-qubit
states can be generalized for mixed states by means of the so-called
\textit{convex roof} \cite{Horodecki:09}
\begin{equation}
\label{c-mixed}
C_N(\rho) = {\rm min} \sum_i p_i \, C_N(\ket{\psi_i}) \, .
\end{equation}
In this latter expression, however, the minimum has to be found with
regard to \textit{all possible} decompositions of $\rho$ into pure
states $\ket{\psi_i}$.

Unfortunately, no solution has been found so far to optimize the
concurrence (\ref{c-mixed}) of a multi-qubit system analytically
\cite{Horodecki:09}, apart from the two-qubit systems
\cite{Wootters:98} and a special case of three qubits
\cite{Lohmayer:06}. Instead of the computationally demanding
optimization of the right-hand side of Eq.~(\ref{c-mixed}), a much
simpler lower bound to this measure has been suggested recently by
Li \etal{} \cite{Li(b):09}
\begin{equation}
\label{low-bound-N}
 C_N(\rho) \geq \tau_N (\rho) \equiv \sqrt{\frac{1}{N}\, \sum_{n=1}^N \,
 \sum_{k=1}^K \,  (C_k^n)^2 } \, .
\end{equation}
This bound is defined in terms of the $N$ ``bi-partite``
concurrences $C^n$ \cite{Ou:08} that correspond to the possible
(bipartite) cuts of the multi-qubit system in which just one of the
qubits is discriminated from the other $N-1$ qubits. For the
separation of the $n$-th qubit, the bipartite concurrence $C^n$ is
given by a sum of $K = 2^{N-2} \,(2^{N-1}-1)$ terms $C_k$ which are
expressed as
\begin{equation}
 \label{concurence}
C_k^{n} = {\rm max} \{ 0, \lambda_k^1 - \lambda_k^2 - \lambda_k^3 -
\lambda_k^4 \} \, ,
\end{equation}
and where the $\lambda_k^m, \, m=1..4$ are the square roots of the
four nonvanishing eigenvalues of the matrix $\rho\,
\tilde{\rho}_k^{n}$, if taken in decreasing order. These
(non-hermitian) matrices $\rho\: \tilde{\rho}_k^{n}$ are formed by
means of the density matrix $\rho$ and its complex conjugate
$\rho^*$, and are further transformed by the operators $\{ S_k^{n} =
L^{n}_k \otimes L_0,\; k = 1,...,K \}$ as: $\tilde{\rho}_k^{n} =
S_k^{n} \rho^\ast S_k^{n}$. In this notation, moreover, $L_0$ is the
(single) generator of the group SO(2), while the $\{ L^{n}_k \}$ are
the $K = 2^{N-2} \,(2^{N-1}-1)$ generators of the group
SO$(2^{N-1})$. We note that the lower bound (\ref{low-bound-N})
reduces to the Wootters concurrence \cite{Wootters:98} for just two
qubits. For details about the explicit construction of the lower
bound (\ref{low-bound-N}) we refer to \cite{Li(b):09}.

Let us display this lower bound (\ref{low-bound-N}) especially for
three-qubits, $\tau_3 (\rho)$, for which we wish later to describe
the entanglement dynamics. For such states, the lower bound $\tau_3
(\rho)$ can be written in terms of the three bipartite concurrences
that correspond to possible cuts of the two qubits from the
remaining one, i.e.
\begin{equation}
\label{low-bound-three}
 \tau_3 (\rho) = \sqrt{\frac{1}{3}\, \sum_{k=1}^6 \,
 (C_k^{12|3})^2  + (C_k^{13|2})^2 + (C_k^{23|1})^2 } \, .
\end{equation}
The bipartite concurrence $C_k^{ab|c}$ (for $a,b,c=1..3$ and $a \neq
b \neq c \neq a$) are obtained as described above with the help of
the operators $\{ S_k^{ab|c} = L^{ab}_k \otimes L^c_0,\; k=1...6
\}$, where $L_0$ is the generator of the group SO(2) which is given
by the second Pauli matrix $\sigma_y = - i \, ( \ket{0}\bra{1} +
\ket{1}\bra{0})$. The (six) generators $L^{ab}_k$ of the group SO(4)
that can be expressed explicitly by means of the totally
antisymmetric Levi-Cevita symbol in four dimensions as
$(L_{kl})_{mn} = - i \, \varepsilon_{klmn}; \; k,l,m,n =1..4$
\cite{Jones:98}.

\subsection{\label{subsec:2.2} Evolution equation for bipartite systems}

Having the explicit expression (\ref{low-bound-N}) for the lower
bound $\tau_N (\rho)$ to the multi-qubit concurrence, we can derive
an evolution equation of this entanglement measure. Since the lower
bound $\tau_N (\rho)$ is given just in terms of the bipartite
concurrences  $C^n$, let us first reconsider the evolution equation
for a bipartite system as suggested by Li \etal{} \cite{Li(d):09}.
Suppose, $\ket{\chi}$ is a pure state of a bipartite system $d_1
\otimes d_2$ with dimension $d_1$ and $d_2$ of the corresponding
subsystems, and the second subsystem undergoes the action of a noisy
channel $\mathcal{S}$. Then, the final state of the system is a
mixed state in general and takes the form $\rho = (1 \otimes
\mathcal{S}) \ket{\chi}\bra{\chi}$. On the other hand, any pure
state $\ket{\chi}$ can be obtained also from the maximally entangled
state $\ket{\phi} = \sum_{i=1}^{d_2} \ket{i} \otimes \ket{i} /
\sqrt{d_2}$ of the bipartite system by $\ket{\chi} = (M \otimes 1)
\ket{\phi}$ \cite{Li(d):09}. In this notation, $M$ denotes a local
(filtering) operator \cite{Gisin:96,Tiersch:09} that acts on the
first subsystem of the maximally entangled state. \cor{Although
local operations, usually associated with projective measurements or
a passage through a channel, cannot change entanglement because of
its nonlocal nature, there is a special case of a stochastic local
operation, a filtering operation, that can be employed to influence
on this nonlocal feature \cite{Tiersch:09}.} Therefore, the final
state $\rho$ of the bipartite system can be expressed as
\begin{equation}
 \label{commutation-1}
 \rho = (1 \otimes \mathcal{S}) \; \left( M \otimes 1 \right) \;
 \ket{\phi}\bra{\phi} \; \left( M^\dagger \otimes 1 \right) \, .
\end{equation}

Since the filtering operator $M$ and the noise $\mathcal{S}$ act
only on either the first or second subsystem, it has been shown in
\cite{Li(d):09} that the reduction of the entanglement of the system
under the action of a noisy channel $\mathcal{S}$ is independent of
the initial state $\ket{\chi}$, and is bounded from above by the
channel's action upon the maximal entangled state $\ket{\phi}$. We
therefore obtain
\begin{equation}
 \label{bipart-evol}
C[(1 \otimes \mathcal{S}) \ket{\chi}\bra{\chi}] \leq \frac{d_2}{2}
C[(1 \otimes \mathcal{S}) \ket{\phi}\bra{\phi}] C[\ket{\chi}] \, ,
\end{equation}
where $C~[..]$ denotes the bipartite concurrence (\ref{concurence}).
If, moreover, the bipartite system consists of a $d_1$-dimensional
and a single-qubit subsystem, and just the qubit is affected by the
noisy channel $\mathcal{S}$, the equal sign applies in inequality
(\ref{bipart-evol}) and we obtain
\begin{equation}
 \label{n-2-evolution}
C[(1 \otimes \mathcal{S}) \ket{\chi}\bra{\chi}] = C[(1 \otimes
\mathcal{S}) \ket{\phi}\bra{\phi}] C[\ket{\chi}] \, .
\end{equation}
That is, the entanglement dynamics of an arbitrary pure state of a
$d_1 \otimes 2$ bipartite system is completely determined by the
channel's action on the maximally entangled state $\ket{\phi}$ of
the bipartite system if the single-qubit subsystem is affected by
the noisy channel $\mathcal{S}$.

We can utilize Eq.~(\ref{n-2-evolution}) to derive next an evolution
equation of the lower bound (\ref{low-bound-three}) to the
three-qubit concurrence, if just one qubit is affected by some noisy
channel. Later, we shall generalize this evolution equation to the
case of a general $N$-qubit system with the same assumption that
just one of its qubits is subjected to a noisy channel.

\subsection{\label{subsec:2.3} Dynamics of the lower bound for initially
                            pure three-qubit states}

Suppose $\ket{\chi}$ is a pure state of a three-qubit system and
just one qubit undergoes the action of a channel $\mathcal{S}$. The
final state of the three-qubit system takes the form $\rho = (1
\otimes 1 \otimes \mathcal{S}) \ket{\chi}\bra{\chi}$, which is
equivalent to the final state of a bipartite $4 \otimes 2$ system
when the second subsystem is subjected to the channel $\mathcal{S}$.
As we mentioned above, any pure state of a bipartite system can be
obtained from the maximally entangled state of the bipartite system
by means of a single local filtering operation $M$ acting on the
first subsystem as $\ket{\chi} = (M \otimes 1) \ket{\phi}$. In
contrast, two local filters $M$ and $M^\prime$ are in general
required to obtain an arbitrary pure three-qubit state $\ket{\chi}$
from a maximally entangled state of three qubits $\ket{\phi}$ by
$\ket{\chi} = (M \otimes M^\prime \otimes 1) \ket{\phi}$. For
three-qubit systems, however, there are two maximally entangled
states which, in the computational basis, are given by
\cite{Duer:00}
\begin{eqnarray}
\label{pure-GHZ}
 & & \ket{\rm GHZ} = \frac{1}{\sqrt{2}} \left( \ket{000} +
\ket{111} \right) \, ,
  \\[0.1cm]
\label{pure-W}
 & & \ket{W} = \frac{1}{\sqrt{3}} \left( \ket{001} +
\ket{010} + \ket{100} \right) \, .
\end{eqnarray}
These two entangled states cannot be obtained from each other by
means of local single-qubit unitary operations \cite{Duer:00} and
give rise to two local unitary inequivalent classes of three-qubit
entangled states, the so-called GHZ- and W-classes.

Although an arbitrary pure three-qubit state $\ket{\chi}$ can be
generated from one of the maximally entangled states by means of
local operations \cite{Duer:00}, we first need to identify the class
of states either (\ref{pure-GHZ}) and (\ref{pure-W}) to which it
belongs to. For an arbitrary (pure or mixed) three-qubit entangled
state, fortunately, this is possible by following the procedure due
to D\"{u}r \etal{} \cite{Duer:00} which is simple and just includes
the computation of the 3-tangle as described in
Ref.~\cite{Coffman:00}. It leads to the distinction that every
entangled three-qubit state $\ket{\chi}$, for which the 3-tangle
vanishes, belong to the W-class and can thus be obtained from the W
state (\ref{pure-W}) by means of local unitary operations. In
contrast, any entangled three-qubit state with nonvanishing 3-tangle
is part of the GHZ-class. For a given pure three-qubit state
$\ket{\chi}$, it is therefore always possible to find proper local
(filtering) operations $M$ and $M^\prime$ so that $\ket{\chi}$ is
obtained from either (\ref{pure-GHZ}) or (\ref{pure-W}) by
$\ket{\chi} = (M \otimes M^\prime \otimes 1) \ket{\phi}$. Moreover,
we have $\ket{\phi} \equiv \ket{\rm GHZ}$ if $\ket{\chi}$ belongs to
the GHZ-class of entanglement, and $\ket{\phi} \equiv \ket{W}$ for
$\ket{\chi}$ being part of the W-class.

To summarize our discussion here, the final state of the three-qubit
system when one of its qubits undergoes the action of a noisy
channel $\mathcal{S}$ is given by
\begin{eqnarray}
 \label{commutation-2}
\rho & = & (1 \otimes 1 \otimes \mathcal{S}) \times \nonumber
 \\[0.1cm]
 & & \left( M \otimes M^\prime \otimes 1 \right) \; \ket{\phi}\bra{\phi}
\; \left( M^\dag \otimes (M^\prime)^\dag \otimes 1 \right) \, .
\end{eqnarray}
where $\ket{\phi}$ is one of the maximally entangled states
$\{\ket{GHZ}, \, \ket{W} \}$. In this equation
(\ref{commutation-2}), the filters $M, \, M^\prime$ and the noise
$\mathcal{S}$ act on different subsystems. This allows us to apply
the evolution equation for bipartite concurrence
(\ref{n-2-evolution}) to a ``bi-partite`` split $12|3$ of the
three-qubit system. We therefore obtain
\begin{eqnarray}
 \label{bipartite-evolution}
C^{12|3}[(1 \otimes 1 \otimes \mathcal{S}) \ket{\chi}\bra{\chi}] & =
& \nonumber
 \\[0.1cm]
& & \hspace*{-2.8cm} C^{12|3}[(1 \otimes 1 \otimes \mathcal{S})
\ket{\phi}\bra{\phi}] \; C^{12|3}[\ket{\chi}] \, ,
\end{eqnarray}
while similar relations can be obtained for the ``bi-partite``
concurrences $C^{13|2}$ and $C^{23|1}$ of the three-qubit system.
Although the Eq.~(\ref{bipartite-evolution}) has similar structure
to the evolution equation for bipartite systems
(\ref{n-2-evolution}), they differ by the maximally entangled state
$\ket{\phi}$ in their right-hand sides: the maximally entangled
state $\ket{\phi} = \sum_{i=1}^{d_2} \ket{i} \otimes \ket{i} /
\sqrt{d_2}$ of the bipartite system is to be substituted in
Eq.~(\ref{n-2-evolution}), while one of the maximally entangled
states (\ref{pure-GHZ})-(\ref{pure-W}) should be used in the right
hand side of Eq.~(\ref{bipartite-evolution}).

Because of the symmetry of the maximally entangled states
(\ref{pure-GHZ})-(\ref{pure-W}) with regard to the qubits
permutation we have a relation
\begin{eqnarray}
 \label{relation}
C^{12|3}[(1 \otimes 1 \otimes \mathcal{S}) \ket{\phi}\bra{\phi}] & =
& C^{13|2}[(1 \otimes \mathcal{S} \otimes 1) \ket{\phi}\bra{\phi}]
\nonumber
 \\[0.1cm]
& & \hspace*{-1.5cm} = C^{23|1}[(\mathcal{S} \otimes 1 \otimes 1)
\ket{\phi}\bra{\phi}] \, ,
\end{eqnarray}
where $\ket{\phi} = \{\ket{GHZ}, \, \ket{W} \}$. From
Eqns.~(\ref{bipartite-evolution}) and (\ref{relation}) it follows
that for an arbitrary pure three-qubit state $\ket{\chi}$ the
evolution of the bipartite concurrence is independent on a bipartite
cut of the three-qubit system, i.e
\begin{eqnarray}
 \label{final-relation}
C^{12|3}[(1 \otimes 1 \otimes \mathcal{S}) \ket{\chi}\bra{\chi}] & =
& C^{13|2}[(1 \otimes \mathcal{S} \otimes 1) \ket{\chi}\bra{\chi}]
\nonumber
 \\[0.1cm]
& & \hspace*{-1.5cm} = C^{23|1}[(\mathcal{S} \otimes 1 \otimes 1)
\ket{\chi}\bra{\chi}] \, .
\end{eqnarray}

Substituting the evolution equation (\ref{bipartite-evolution}) into
definition of the lower bound (\ref{low-bound-three}) and taking
into account relation (\ref{relation}), we finally obtain an
evolution equation of the lower bound for three-qubit concurrence
\begin{equation}
 \label{tau}
\tau_3 [(1 \otimes 1 \otimes \mathcal{S}) \ket{\chi}\bra{\chi}] =
\tau_3 [(1 \otimes 1 \otimes \mathcal{S}) \ket{\phi}\bra{\phi}] \:
\tau_3 [\ket{\chi}] \, ,
\end{equation}
where $\tau_3~[..]$ is defined in Eq.~(\ref{low-bound-three}). The
entanglement dynamics of an arbitrary pure state $\ket{\chi}$ of a
three-qubit system, when one of its qubits undergoes the action of
an arbitrary noisy channel $\mathcal{S}$, is subjected to the
dynamics of one of the maximally entangled states $\ket{\phi} = \{
\ket{GHZ}, \, \ket{W} \}$. The choice between the maximally
entangled states should be done after determining the entanglement
class of the given state $\ket{\chi}$ following the procedure in
Ref.~\cite{Duer:00} and briefly discussed above. We note, that due
to Eq.~(\ref{final-relation}) the entanglement dynamics of a pure
three-qubit state $\ket{\chi}$ is independent of which of the qubits
is affected by the noise. In fact, this equation
(\ref{final-relation}) significantly simplifies the calculation of
the lower bound (\ref{low-bound-three}). It is sufficient to compute
just one bipartite concurrence in definition (\ref{low-bound-three})
of the lower bound, for example $C^{12|3}[\ket{\chi}]$, while the
bipartite concurrences $C^{13|2}[\ket{\chi}]$ and
$C^{23|1}[\ket{\chi}]$ are equal to it due to
Eq.~(\ref{final-relation}).

\subsection{\label{subsec:2.4} Remarks on the evolution of pure $N$-qubit
            and mixed three-qubit states}

It is desirable, of course, to generalize the evolution equation
(\ref{tau}) of the lower bound to the three-qubit concurrence also
to $N$-qubit states, if just one of the qubits is affected by a
noisy channel $\mathcal{S}$. In contrast to the classification of
the three-qubit states, however, it is not known until now how many
and which entanglement classes eventually exist for qubit systems
with $N > 4$, while some classification is available for $N=4$
\cite{Verstraete:02,Li:09}. It is therefore not directly possible to
generalize Eq.~(\ref{tau}) for arbitrary pure states of $N$ qubits.
Nevertheless, some entanglement classes are known also for general
pure N-qubit states, such as the GHZ- and W-class. If a given (pure)
N-qubit state $\ket{\chi}$ belongs to the GHZ- or W-class, the
evolution equation (\ref{tau}) of the lower bound can be extended to
\begin{eqnarray}
 \label{tau-N}
 & & \tau_N [(1^{\otimes N-1} \otimes \mathcal{S})
     \ket{\chi}\bra{\chi}] =
 \nonumber
 \\[0.1cm]
 & & \hspace*{1.1cm}
     \tau_N [(1^{\otimes N-1} \otimes \mathcal{S})
     \ket{\phi}\bra{\phi}] \: \tau_N [\ket{\chi}] \, ,
\end{eqnarray}
where $\ket{\phi}$ denotes the corresponding maximal entangled
$N$-qubit state \cite{Wu:06}
\begin{eqnarray}
\label{GHZ-N}
 \ket{\rm GHZ}_N = \frac{1}{\sqrt{2}} \left(
\ket{0}^{\otimes N} + \ket{1}^{\otimes N} \right) \, ,
\\[0.1cm]
\label{W-N}
 \ket{W}_N = \frac{1}{\sqrt{N}} \left( \ket{10,...,0} +
 \ket{01,...,0} + \ket{00,...,1} \right) \, .
\end{eqnarray}

We can further analyze the lower bound (\ref{low-bound-three}) to
the three-qubit concurrence in order to understand the entanglement
evolution in those cases where one starts already with an initially
mixed state $\rho_0$. Exploiting the convexity of the lower bound
(\ref{tau}) as a valid entanglement measure, we have $ \tau_3 [ \,
(1 \otimes 1 \otimes \mathcal{S}) \, \rho_0 \,]  \,=\,
  \tau_3 [ \, \sum_i \, p_i (1 \otimes 1 \otimes \mathcal{S})
              \ket{\psi}_i \bra{\psi}_i \, ]                   \,\leq\,
  \sum_i \, p_i \tau_3 [ \, (1 \otimes 1 \otimes \mathcal{S})
                         \ket{\psi}_i \bra{\psi}_i \, ] $.
Making use of this inequality in Eq.~(\ref{tau}), we obtain
\begin{equation}
 \label{tau-mixed}
\tau_3 [(1 \otimes 1 \otimes \mathcal{S}) \rho_0] \leq \tau_3 [(1
\otimes 1 \otimes \mathcal{S}) \ket{\phi}\bra{\phi}] \: \tau_3
[\rho_0] \,
\end{equation}
for the evolution of the lower bound and for an initially mixed
state. As before, we assume here that just one of the qubits is
affected by the noisy channel $\mathcal{S}$. Moreover, the
inequality (\ref{tau-mixed}) can be generalized for local two- and
three-sided channels, i.e.~to cases in which two or even all three
qubits are affected by some local noise. For example, for a local
two-sided channel $ \mathcal{S}_1 \otimes \mathcal{S}_2 \otimes 1 \,
= \,
 (\mathcal{S}_1 \otimes 1 \otimes 1) \, (1 \otimes \mathcal{S}_2 \otimes 1)$
we find
\begin{eqnarray}
\label{tau-three-sided}
   \tau_3 [(\mathcal{S}_1 \otimes \mathcal{S}_2 \otimes 1) \rho_0]
   &\leq &
   \tau_3 [(\mathcal{S}_1 \otimes 1 \otimes 1) \ket{\phi}\bra{\phi}] \:
   \nonumber
   \\[0.1cm]
   & & \hspace*{-1.6cm} \times\,
   \tau_3 [(1 \otimes \mathcal{S}_2 \otimes 1)
   \ket{\phi}\bra{\phi}] \: \tau_3 [\rho_0] \, .
\end{eqnarray}
It is this particular form of Eq.~(\ref{tau-three-sided}) that gives
rise to a sufficient criterion for finite-time disentanglement of
arbitrary initial states being subjected to local multi-sided
channels \cite{Konrad:08}.

\section{\label{sec:3} Dynamics of the initially mixed state}

The three evolution equations (\ref{tau}), (\ref{tau-mixed}) and
(\ref{tau-three-sided}) provide us with a powerful tool in order to
describe the time-dependent entanglement dynamics of a three-qubit
system under local noise. \cor{Although the assumption was made that
only one qubit is affected by a noisy channel [apart from
Eq.~(\ref{tau-three-sided})], nothing more need to be known
explicitly about the time evolution of the system's state.} In the
present section, we make use of Eq.~(\ref{tau-mixed}) in analyzing
the entanglement dynamics of the initially mixed state
(\ref{the-mixed-state}) for a variety of particular single-qubit
noise models. Using the definition (\ref{pure-GHZ})-(\ref{pure-W})
of the $\ket{\rm GHZ}$ and $\ket{W}$ in Eq.~(\ref{the-mixed-state}),
we can compute the various parts of inequality (\ref{tau-mixed})
analytically when one qubit undergoes the action of noisy channels.

Indeed, there are several reasons for studying the entanglement
evolution of the mixed state (\ref{the-mixed-state}). For this
state, first of all, an analytical expression is known for the
convex roof to the concurrence (\ref{c-mixed}) \cite{Lohmayer:06}.
This enables one to compare the time-dependent lower bound from the
evolution equation (\ref{tau-mixed}) with the behavior of the convex
roof as deduced from the state dynamics under the influence of a
certain noise model. Second, the mixed state density matrix
(\ref{the-mixed-state}) has simply rank two. As we have recently
shown in particular examples, for rank-2 density matrices the lower
bound (\ref{low-bound-three}) coincides with the convex roof
\cite{Siomau:10}. Third, it will be quite easy to compare also the
speed of `disentanglement' for an (initially) pure GHZ state
$\rho(t,p=1)$ and the pure W state $\rho(t,p=0)$ under decoherence.
We also note that the values for the lower bound for these pure
states are related to each other through the ratio
\begin{equation}
 \label{GHZvsW}
 \frac{\tau_3(\rho(t=0,p=1))}{\tau_3(\rho(t=0,p=0))} =
 \frac{\tau_3(\ket{GHZ})}{\tau_3(\ket{W})}  = \frac{3}{2\sqrt{2}} \,
 ,
\end{equation}
a result that was obtained in Ref.~\cite{Carvalho:04} by means of a
lower bound to the concurrence, different from definition
(\ref{low-bound-three}).

In the evolution equation (\ref{tau-mixed}) we need to obtain the
final state of the three-qubit system $\rho_{\rm fin}$ under the
influence of a local noise $\mathcal{S}$, i.e. $\rho_{\rm fin} = (1
\otimes 1 \otimes \mathcal{S}) \, \rho_{\rm ini}$, where
$\rho_{\rm\, ini}= \{\rho_0, \, \ket{\phi}\bra{\phi} \}$. Following
quantum operation formalism \cite{Nielsen:00}, the final state can
be obtained with the help of (Kraus) operators \cite{Kraus:83} as
\begin{equation}
 \label{sum-represent}
 \rho_{\rm fin} = \sum_i K_i \, \rho_{\rm ini} \, K_i^\dag \, .
\end{equation}
and the condition $\sum_i K_i^\dag \, K_i \leq I$ is fulfilled. The
Kraus operators are well known for various single-qubit noise
models; in the next section, we consider two frequently used noises,
namely the phase and the amplitude damping channels which are
associated with relevant physical processes.

\subsection{\label{subsec:3.1} Phase and amplitude damping channels}

\begin{figure}
\begin{center}
\includegraphics[scale=0.7]{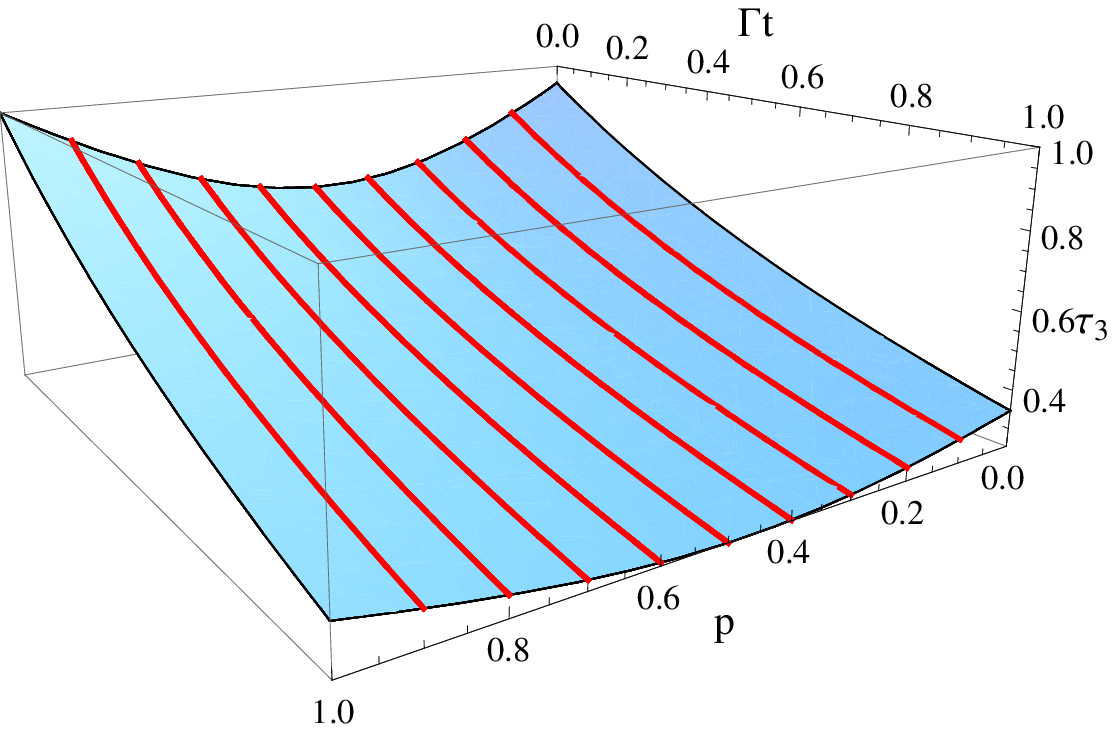}
\caption{(Color online) Evolution of the lower bound $\tau_3(\rho)$
for initially mixed state (\ref{the-mixed-state}) if affected by the
phase damping channel. While the blue surface shows the lhs of
inequality (\ref{tau-mixed}), the red lines represents its rhs.}
 \label{fig-1}
\includegraphics[scale=0.7]{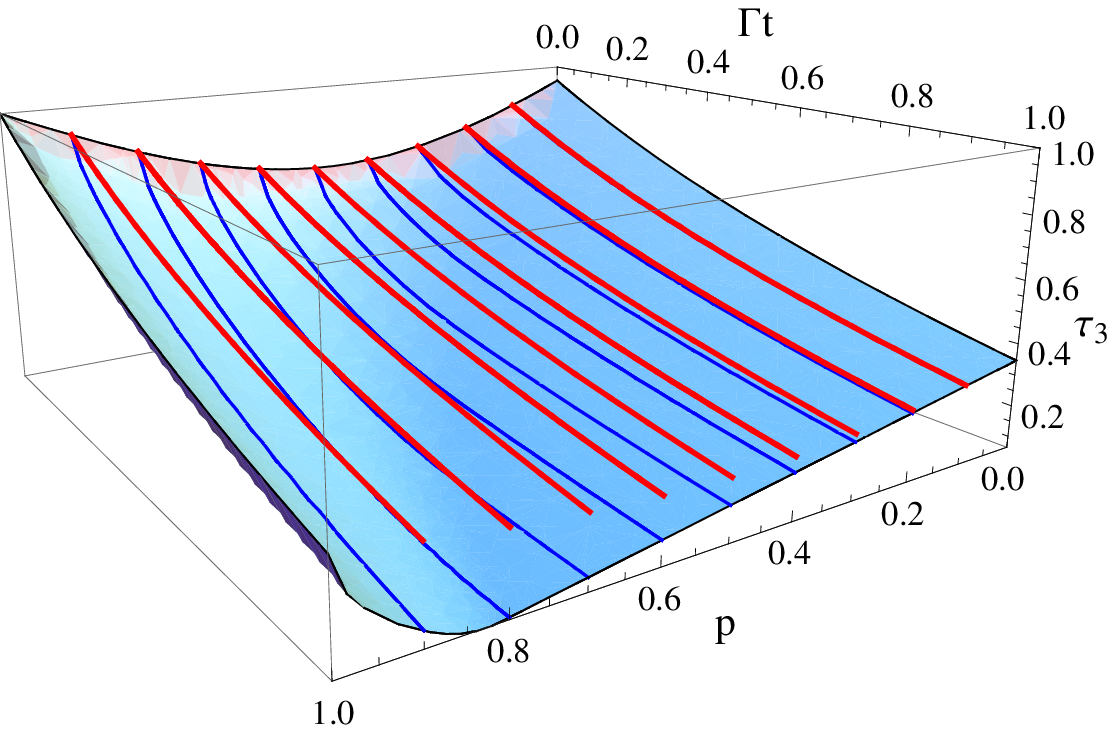}
\caption{(Color online) The same as in Fig.~1 but for the amplitude
damping channel.}
 \label{fig-2}
\end{center}
\end{figure}

Let us start the discussion with a three-qubit system which is
prepared initially in the state (\ref{the-mixed-state}) and for
which just one qubits undergoes the action of the phase damping
channel. A phase damping describes for instance a diffusive
scattering interaction of the qubit with its environment and is
known to result into a loss of phase coherence information
\cite{Nielsen:00}. A possible representation of the phase damping in
terms of time-dependent (Kraus) operators is given by
\cite{Nielsen:00}
\begin{equation}
 \label{phase-damping}
K_1^{\rm pd} = \left( \begin{array}{cc} e^{- \Gamma t} & 0 \\ 0 & 1
\end{array} \right) , \quad K_2^{\rm pd} = \left( \begin{array}{cc}
\sqrt{1-e^{- 2\Gamma t}} & 0 \\ 0 & 0 \end{array} \right) \, .
\end{equation}
where $\Gamma$ denotes a coupling constant. For this noise model,
Fig.~\ref{fig-1} displays the time-dependent evolution of the lower
bound $\tau_3$ for the initially prepared state
(\ref{the-mixed-state}) for different parameters $p$ of the mixed
state and at different times of the system-channel coupling. In this
figure, the blue surface displays the left-hand side (lhs) of the
inequality (\ref{tau-mixed}), while the red lines shows the
corresponding right-hand side. For all parameters $0 \le p \le 1$ of
the mixed state, the lower bound $\tau_3$ decays exponentially and
vanishes only asymptotically for $t \rightarrow \infty$. For the
phase damping channel, moreover, the lhs and rhs of
(\ref{tau-mixed}) are always equal for an arbitrary parameter $p$
and for all times $t$.

If the same system is affected by a (local) amplitude damping
channel, which describes the dissipative coupling of a qubit to a
thermal reservoir in the zero-temperature limit \cite{Nielsen:00},
the operator elements in Eq.~(\ref{sum-represent}) are given by
\begin{equation}
 \label{amplitude-damping}
   K_1^{\rm ad} = \left( \begin{array}{cc} \cor{1} & 0 \\ 0 & \cor{e^{- \Gamma t}}
                         \end{array} \right) , \quad
   K_2^{\rm ad} = \left( \begin{array}{cc} 0 & \sqrt{1-e^{- 2\Gamma t}} \\
                                           0 & 0 \end{array} \right) \, .
\end{equation}
For such an amplitude damping, the time evolution for the lower
bound $\tau_3$ differs from the corresponding dynamics in the phase
damping channel as seen from Fig.~\ref{fig-2}. In this amplitude
damping model, in particular, the (blue) surface for the lhs of
(\ref{tau-mixed}) differs significantly from the rhs of this
inequality for some values of the parameter $p$.

\subsection{\label{subsec:3.2} Sudden death of three-qubit entanglement}

\begin{figure}
\begin{center}
\includegraphics[scale=0.7]{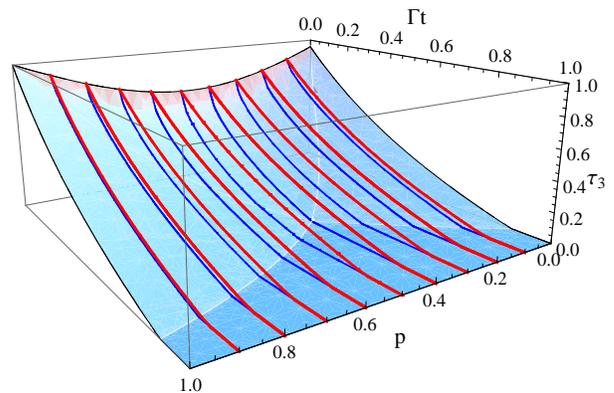}
\caption{(Color online) The same as in Fig.~1 but for the
generalized amplitude damping channel. The sudden death of
entanglement is clearly seen for all $0 \le p \le 1$.}
 \label{fig-3}
\end{center}
\end{figure}

Based on violation of additivity in the case of entanglement decay
in a multi-qubit system coupled to two independent weak noises,
entanglement sudden death \cite{Yu:06} reveals a practically
important aspect of time-dependent entanglement evolution. It is
important to verify whether such a phenomenon can be predicted with
the suggested evolution equation of the lower bound
(\ref{tau-mixed}). Let a three-qubit system be prepared in the state
(\ref{the-mixed-state}) and one of the qubits is subjected to the
generalized amplitude damping channel. This noisy channel can be
viewed as a `superposition' of two independent amplitude damping
channels acting on a qubit and can be expressed by four Kraus
operators $K_1^{\rm gad} = \frac{1}{2} K_1^{\rm ad}, \, K_2^{\rm
gad} = \frac{1}{2} K_2^{\rm ad}$ and
\begin{equation}
 \label{gen-ad}
K_3^{\rm gad} = \frac{1}{2} \left( \begin{array}{cc} \cor{e^{-
\Gamma t}} & 0 \\ 0 & \cor{1} \end{array} \right) , \quad K_4^{\rm
gad} = \frac{1}{2} \left( K_2^{\rm ad} \right)^\dag \, ,
\end{equation}
where $K_1^{\rm ad}$ and $K_1^{\rm ad}$ are defined by
Eq.~(\ref{amplitude-damping}). The evolution of the lower bound
$\tau_3$ for the three-qubit state (\ref{the-mixed-state}) is shown
in Fig.~\ref{fig-3}. Although the lhs and the rhs of the inequality
(\ref{tau-mixed}) differ for some parameters $p$ of the initial
state $\rho(p)$, they both vanish in a finite time for all values
$p$.

\section{\label{sec:4} Conclusion}

Unlike the evolution equations of concurrence for two qubits
\cite{Konrad:08} and for a bipartite system \cite{Li(d):09}, we have
presented an evolution equation of the lower bound for multi-qubit
concurrence (\ref{tau}) for an initially pure state of the system
and when just one qubit is affected by local noise. This evolution
equation (\ref{tau}) has been also extended to the cases of
initially mixed states (\ref{tau-mixed}) and the action of
many-sided noisy channels (\ref{tau-three-sided}). In addition, the
evolution equation of the lower bound for initially mixed states
(\ref{tau-mixed}) has been employed especially to show entanglement
dynamics of the mixed state (\ref{the-mixed-state}) when just one
qubit is affected by the phase, the amplitude or the generalized
amplitude damping channel [cf.~Figures~\ref{fig-1}--\ref{fig-3}].

Of course, the lower bound (\ref{low-bound-N}) is only an
approximation to the convex roof for multi-qubit concurrence
(\ref{c-mixed}) for which an analytically computable expression is
currently unavailable. Nevertheless, we have recently shown that
this lower bound coincided with the convex roof for some scenarios
of entanglement dynamics \cite{Siomau:10}.  A more detailed analysis
of the accuracy of the lower bound approximation is presently under
work.

\begin{acknowledgments}
M.S. thanks Andreas Osterloh for a discussion and Sean McConnell for
his careful reading of the manuscript and useful feedback. This work
was supported by the DFG under the project No. FR 1251/13.
\end{acknowledgments}

\end{document}